\documentstyle[epsf,amsfonts,amssymb,eqsecnum,aps,prb,twocolumn]{revtex}
\begin{document}
\draft
\wideabs{
\title{Colloid aggregation induced by oppositely charged polyions}
\author{Ludger Harnau$^{1,2}$
 and Jean-Pierre Hansen$^{3}$
}
\address{
         $^{1}$Max-Planck-Institut f\"ur Metallforschung,  
         Heisenbergstr.\ 1, D-70569 Stuttgart (Germany) 
         \\
         $^{2}$Institut f\"ur Theoretische und Angewandte Physik, 
         Universit\"at Stuttgart, 
         Pfaffenwaldring 57, 
         D-70569 Stuttgart (Germany) 
         \\ 
         $^{3}$Department of Chemistry, University of Cambridge, Lensfield Road,
         Cambridge CB2 1EW (UK)
} 
\date{\today}
\maketitle
\sloppy
\begin{abstract}
The "polymer reference interaction site model" (PRISM) integral equation 
formalism is used to determine the pair structure of binary colloidal 
dispersions involving large and small polyions of opposite charge. Two 
examples of such bidisperse suspensions are considered in detail, namely 
mixtures of charged spherical colloids and oppositely charged polyelectrolyte
chains, and binary mixtures of oppositely charged large and small clay 
platelets. In both cases clear evidence is found for aggregation of the 
larger particles induced by the polyionic counterions, signalled by a strong 
enhancement of long wavelength concentration fluctuations.
\end{abstract}
}
\narrowtext

\section{Introduction}  
In metallurgy it has been known for several millenia that mechanical 
properties of many metals can be considerably improved by alloying two 
or several elements. Similar improvements of material properties of "soft 
matter" can be achieved by mixing two or more macromolecular components
in fluid solutions or dispersions. One of the oldest known examples is the 
stabilization of colloidal suspensions by the addition of adsorbing 
polymer, which was already used in ancient Egypt for the preparation 
of inks. On the contrary, flocculation of colloids can be driven by the 
addition of non-adsorbing polymer via the depletion mechanism \cite{loui:01}.
More recent is the realization of the key role of Coulombic interactions 
in determining and improving material or physiological properties of 
complex colloidal and biomolecular systems, and their exploitation 
in numerous technological and biomedical applications. Widely studied 
systems include suspensions of charged colloids and polyelectrolytes 
\cite{shar:97,mate:99}, polymer-clay nanocomposites \cite{gian:99}
or DNA bundles and DNA-lipid complexes \cite{gelb:00}. For example, 
even small polyelectrolyte additives can have a substantial impact 
on the aggregation and kinetic stability of charged colloidal 
particles \cite{mate:99}.
On the other hand, just a small fraction of clay fillers doubles the 
tensile strength of a polymer matrix \cite{koji:93}, while highly 
asymmetric mixtures of charged clay platelets are used to improve 
rheological properties of drilling fluids \cite{schl}.

To design complex mixtures with optimal properties, it is essential 
to gain a quantitative understanding of correlations and effective 
interactions between various macromolecular species. This can be 
achieved using the well-established techniques of liquid-state theory 
\cite{hans:86}. The PRISM ("polymer reference interaction site model") 
theory, originally designed for the study of polymer solutions and melts 
\cite{schw:97}, has been recently extended to investigate various 
charged and uncharged colloidal systems \cite{harn:00,harn:01}, as 
well as mixtures of neutral spherical colloids and polymers 
\cite{yeth:92,khal:97,chat:98,fuch:00} and like-charged spherical colloids and 
polyelectrolytes \cite{ferr:00}.

This paper focuses on the structure and stability of dispersions of 
charged colloidal particles in the presence of oppositely charged polyions, 
which play the role of mesoscopic counterions. There is a considerable
literature on the widely studied case of microscopic counterions and 
the resulting effective interactions between the highly charged 
colloidal particles 
\cite{hans:00}. In particular, it is now well established that divalent 
counterions can lead to overscreening of the Coulombic interaction 
between spherical or rod-like colloidal particles, 
which results in a correlation-induced, short-range attraction between
the latter \cite{gron:98,alla:98,gron:97}. The case where the counterions 
are mesoscopic or macromolecular polyions has received much less attention 
so far, except for the simplest case of polyelectolyte chains confined 
between two infinite parallel plates carrying charges of opposite sign 
to the polyelectrolyte, where overcompensation of the charge on the 
platelets, leading to effective attraction between the latter, has been 
predicted \cite{sjos:96,boru:98}.

Two systems involving mesoscopic counterions are examined in the present 
paper, namely a) mixtures of highly charged spherical colloids and 
oppositely charged polyelectrolyte chains and b) binary mixtures of 
oppositely charged platelets of widely different diameters. The key 
problem to be addressed is that of structural diagnostics of a possible
aggregation of the larger particles induced by the smaller particles, 
over a range of physical conditions. This goal is achieved by 
determining  pair correlation functions and partial structure 
factors within the framework of PRISM, which is adapted in this paper
to the systems in hand.

\section{Interaction site models}  
The two systems under investigation are aqueous dispersions or solutions, 
but in view of the mesoscopic scale of the particles, the solvent will 
be modelled as a structureless dielectric continuum providing a macroscopic
permittivity $\varepsilon$. Any microscopic counterions or ions from added 
electrolyte will be considered at the linear response (or Debye-H\"uckel) 
level, i.e., they will screen the electrostatic potential due to the 
interaction sites on the colloids or polyions on a scale given by the 
usual Debye screening length $\lambda_D$ \cite{kutt:00}. The underlying 
Born-Oppenheimer-like assumption entails that the charge distribution on 
the mesoscopic particles does not contribute to screening \cite{warr:00}.
Since sites with charge of opposite sign are involved, the site-site 
interaction must contain a short-range repulsion, which will be modelled
by hard sphere (HS) cores of diameter $d_i$. The interaction potential 
between sites on particles of species $i$ and $j$, carrying the charges 
$z_ie$ and $z_je$, will hence be of the generic form:
\begin{equation}  \label{eq1}
u_{ij}(r)=
\left\{ 
\begin{array}{ll}
\infty\,,\hspace{1cm}&r\le\frac{1}{2}(d_i+d_j)\,,
\nonumber
\\\frac{\displaystyle z_iz_je^2}{\displaystyle\varepsilon r}
\exp(-\kappa_Dr)\,,\hspace{0.1cm}&r>\frac{1}{2}(d_i+d_j)\,,\end{array}\right.
\end{equation}
where $\kappa_D=\lambda_D^{-1}$.
On the polyelectrolyte, $N_p$ charged sites are distributed at regular 
intervals along the linear macromolecular backbone, each site being associated
with a monomer or segment of diameter $d_p$. The clay platelets will be 
modelled as circular discs of radius $R$, carrying interaction sites 
uniformly distributed on the surface; the core diameter associated with 
these sites accounts for the finite thickness of the platelets. As regards 
the spherical colloids, two models may be considered. In model A, $N_s$ 
charged sites are distributed over the spherical surface of radius $R$.
For a sufficiently large density $N_s/(4\pi R^2)$ of the surface sites, 
the hard cores of diameter $d_s$ associated with each site will ensure 
that the spheres are impenetrable. In model B, the impenetrability 
of the spherical colloids is ensured by an uncharged central ($c$) 
site, with a core radius identical to the particle radius, while the 
electrostatic interactions are carried by $N_s$ surface sites distributed 
on a spherical shell of radius $R-d_s/2$. Each surface site carries a charge 
equal to the total charge on the colloid divided by the number of sites.
Invoking Gauss' theorem, it might be argued that this distribution should 
be equivalent to a single site carrying the full colloidal charge at the 
centre of the sphere. This equivalence would hold within an exact theory, 
but since an approximate closure relation is to be used, as discussed 
in section III, more accurate results may be expected when interaction 
sites with lower charges are used, such that the Coulomb coupling is 
weaker.

Within the PRISM formalism to be discussed in the next section, all 
interaction sites of the same type (i.e., $p$, $c$, or $s$) are assumed 
to be equivalent, leading to a considerable simplification over the 
original RISM \cite{chan:82}. This assumption is obviously exact for 
spherical particles with surface sites only (model A), but is an 
approximation for polyelectrolytes of finite length (since end 
effects are neglected), and even more so for platelets, where edge 
effects are not taken into consideration. However previous experience 
with neutral or charged rods \cite{harn:00} and platelets \cite{harn:01}
shows that the basic PRISM assumption does lead to sensible results for 
the pair structure of these rigid particles.

\section{Multicomponent PRISM}
Consider a multicomponent system involving $\nu$ species of particles 
with number densities $\rho_\alpha$ ($1\le \alpha \le \nu$), each 
particle of a species involving $n_\alpha$ classes of exactly or 
approximately equivalent interaction sites. The total number of 
distinct classes of sites in the mixture is 
\begin{equation} \label{eq2}
\sum\limits_{\alpha=1}^\nu n_\alpha=N\,.
\end{equation}
It proves convenient to assign an index $i$ ($1\le i \le N$) to 
each class of sites, and to order them such that classes 
$1\le i \le n_1$ belong to particles of species $1$, classes 
$n_1+1 \le i \le n_1+n_2$ belong to species $2$, ... , and classes 
$N-n_\nu \le i \le N$ belong to species $\nu$. Each class $i$ contains 
$N_i$ equivalent interaction sites, and if the positions of sites 
belonging to classes $i$ and $j$ are designated by 
${\bf r}_k\equiv {\bf r}^{(i)}_k$, and  ${\bf r}_l\equiv {\bf r}^{(j)}_l$
($1 \le k \le N_i\,;\,\, 1 \le l \le N_j$), then one may define the usual 
intramolecular structure factors, or form factors
\begin{equation} \label{eq3}
\omega_{ij}({\bf q})=\frac{1}{\sqrt{N_iN_j}}\sum\limits_{k=1}^{N_i}
\sum\limits_{l=1}^{N_j}\left\langle 
e^{i{\bf  q} \cdot ({\bf r}_k-{\bf r}_l)}\right\rangle
\end{equation}
provided $i$ and $j$ sites belong to the same particle, while 
$\omega_{ij}({\bf q})\equiv 0$ otherwise. The $\omega_{ij}({\bf q})$
are hence elements of a $N$ x $N$ box-diagonal matrix 
\mbox{\boldmath$\omega$}$({\bf q})$. 
In the present paper {\em rigid} platelets 
and spherical colloids are considered, so that 
\begin{equation} \label{eq4}
\omega_{ij}(q)=\delta_{ij}+
\frac{1}{\sqrt{N_iN_j}}\sum\limits_{k=1}^{N_i}
\sum\limits_{l=1}^{N_j\,\,'}
\frac{\sin(qr_{kl})}{qr_{kl}}\,,
\end{equation}
where $r_{kl}=|{\bf r}_k-{\bf r}_l|$, $q=|{\bf q}|$, and
the prime signifies that 
the ''self'' term $k=l$ is to be left out for the diagonal elements 
$i=j$. PRISM is based on the assumption that all direct correlation 
functions between sites on pairs of different particles are identical 
if these sites belong to the same class on each of the two particles. 
This leaves a total of $N(N+1)/2$ independent direct correlation 
functions $c_{ij}(q)$. Corresponding total correlation functions 
$h_{ij}(q)$ are defined by averaging over the $N_iN_j$ correlation 
functions between all pairs of equivalent sites belonging to classes 
$i$ and $j$ on two particles of the same or different species. The 
$N$ x $N$ matrices of $c_{ij}(q)$ or $h_{ij}(q)$ are 
symmetric, but it proves convenient to introduce auxiliary non-symmetric 
matrices ${\bf H}(q)$, ${\bf C}(q)$ and ${\bf W}(q)$,
with elements
\begin{eqnarray} \label{eq5}
H_{ij}(q)&=&\rho_iN_iN_jh_{ij}(q)\,,\hspace{0.1cm}
C_{ij}(q)=\rho_ic_{ij}(q)\,,\,
\\W_{ij}(q)&=&\sqrt{N_iN_j}\omega_{ij}(q)\,,
\end{eqnarray} \label{eq5a}
where $\rho_i$ is the number density of that species of particles 
to which sites of category $i$ belong. With these conventions, the 
set of $N(N+1)/2$ Ornstein-Zernike (OZ) relations can be cast in the 
compact form \cite{harn:00,chan:82,shew:99}
\begin{equation} \label{eq6}
{\bf H}(q)={\bf W}(q){\bf C}(q)[{\bf W}(q)+{\bf H}(q)]
\end{equation}
which may be solved for ${\bf H}(q)$ according to:
\begin{equation} \label{eq7}
{\bf H}(q)=\left([{\bf I}-{\bf W}(q){\bf C}(q)]^{-1}
-{\bf I}\right){\bf W}(q)\,.
\end{equation}
The set of $N(N+1)/2$ independent OZ relations must be supplemented 
by as many closure relations between each pair of total and direct 
correlation functions. For intersite distances $r>d_{ij}=(d_i+d_j)/2$, 
we adopt the partially linearized Laria-Wu-Chandler closure
\cite{lari:91,yeth:93}:
\begin{eqnarray}  \label{eq8a}
&&\frac{h_{ij}(r)+1}{h_{ij}^0(r)+1}
\exp[h_{ij}^0(r)-\omega_{ii}*c_{ij}^0*\omega_{jj}(r)]\nonumber
\\&&=
\left\{
\begin{array}{ll}
\exp[\chi_{ij}(r)]\,,\hspace{0.3cm}&\chi_{ij}(r)\le 0\\
1+\chi_{ij}(r)\,,\hspace{0.3cm}&\chi_{ij}(r)>0\,,
\end{array}\right.
\end{eqnarray}
with
\begin{equation}  \label{eq8b}
\chi_{ij}(r)=h_{ij}(r)-
\omega_{ii}*(\beta u_{ij}+c_{ij})*\omega_{jj}(r)\,,
\end{equation}
where $\beta=1/(k_BT)$ is the inverse temperature. The superscript $0$ 
refers to the underlying reference model with hard core interactions 
only (i.e., without the screened Coulomb interaction in Eq.~(\ref{eq1})),
while the asterisk $*$ denotes a convolution product. The reference 
site-site correlation functions $c_{ij}^0(r)$ and $h_{ij}^0(r)$ 
are obtained by solving the set of OZ equations (\ref{eq6}) with 
the atomic Percus-Yevick closure for hard spheres, i.e. 
$h_{ij}^0(r)=-1$, $r\le d_{ij}$; $c_{ij}^0(r)=0$, $r>d_{ij}$ \cite{hans:86}.
The closure (\ref{eq8a}) takes the long-range asymptotics of the 
direct correlation functions $c_{ij}(r)$, in terms of the molecular 
hypernetted chain (HNC) approximation \cite{hans:86,chan:82}, into 
account, but prevents the exponential increase of the total correlation 
functions $h_{ij}(r)$ for distances corresponding to strong attraction 
between oppositely charged sites \cite{kova:99}. By construction, the 
total correlation functions and their first derivatives are continuous 
when $\chi_{ij}(r)=0$.

The closed set of equations (\ref{eq7}) - (\ref{eq8b}) 
are solved numerically by a standard iterative procedure to obtain 
pair distribution functions $g_{ij}(r)=1+h_{ij}(r)$, and partial 
structure factors: 
\begin{equation}  \label{eq9}
S_{ij}(q)=
\omega_{ij}(q)\delta_{ij}+\rho_i h_{ij}(q)
\,.
\end{equation}

\section{Colloid-polyelectrolyte mixtures}
In this binary mixture ($\nu =2$), species $1$, referred to with the 
index $p$, are semiflexible polyelectrolyte chains with $N_p$ 
equivalent segments carrying each a charge $z_pe=Z_pe/N_p$
(where $Z_pe$ is the total charge of each chain). The form 
factor of the semiflexible chain is \cite{harn:96}:
\begin{eqnarray}  \label{eq10}
\omega_p(q)&=&1+\frac{2}{N_p}\sum\limits_{n=1}^{N_p-1}(N_p-n)
\nonumber
\\&&\times \exp\left[-\frac{q^2}{3}[nd_pl_p-
l_p^2(1-e^{-nd_p/l_p})]\right]\,,
\end{eqnarray}
where $d_p$ is the segment length and $l_p$ is the persistence length. 
The spherical colloids of radius $R_c$ (species $2$), carry two categories 
of sites: $N_s$ rigidly arranged surface sites of diameter $d_s$ 
carrying an electric charge $z_se=Z_ce/N_s$ (where $Z_ce$ is the total 
charge of each colloidal particle), and one central site, labelled 
with the index $c$, which is non-interacting (model A), or the centre 
of a hard sphere repulsion of radius $R_c$ (model B). In this paper 
only calculations based on model A will be presented. The corresponding 
form factors are $\omega_{cc}(q)=1$,
\begin{eqnarray}  \label{eq11}
\omega_{ss}(q)&=&1+N_s\left(\frac{\sin(qR_c)}{qR_c}\right)^2\,,
\\\omega_{cs}(q)&=&\sqrt{N_s}\,\frac{\sin(qR_c)}{qR_c}\,,  \label{eq12}
\end{eqnarray}
where the double sums in Eq.~(\ref{eq4}) have been replaced by double 
integrals corresponding to a continuous distribution of interaction 
sites on the surface of the sphere of radius $R_c$. Since this 
corresponds formally to the limit $N_s\to \infty$, the two form factors 
(\ref{eq11}) and (\ref{eq12}) are approximately related by 
$\omega_{ss}(q)=\omega^2_{cs}(q)$. In practical calculations $N_s=60$ 
was chosen. If no microscopic co- and counterions are present, charge 
neutrality relates the polyelectrolyte and colloid density via:
\begin{equation}  \label{eq13}
Z_c\rho_c+Z_p\rho_p=0\,.
\end{equation}
In the absence of microscopic ions, $\kappa_D=0$, and the potentials 
$u_{ij}(r)$ in Eq.~(\ref{eq1}) reduce to bare Coulomb potentials. 
With three classes of sites ($p$, $s$ and $c$), the matrix ${\bf W}(q)$
is given by:
\begin{equation}
{\bf W}(q)=
\left(
\begin{array}{lll}       \label{eq15}
\omega_{pp}(q)&0&0\\
0&\omega_{ss}(q)&\omega_{cs}(q)\\
0&\omega_{cs}(q)&\omega_{cc}(q)
\end{array}
\right)\,.
\end{equation}

There are a priori 6 independent direct and total correlation functions 
$c_{ij}(q)$ and $h_{ij}(q)$ ($i,j=p, s$ or $c$). However, independently 
of any specific closure relation, the OZ equations lead directly to 
the following three relations between the total correlation functions:

\begin{eqnarray} 
h_{ss}(q)&=&\omega^2_{cs}(q)h_{cc}(q)\,, \label{eq16a}
\\h_{cs}(q)&=&\omega_{cs}(q)h_{cc}(q)\,, \label{eq16b}
\\h_{ps}(q)&=&\omega_{cs}(q)h_{pc}(q)\,. \label{eq16c}
\end{eqnarray}
These relations reflect the fact that the ''free-rotation'' approximation 
\cite{hans:86} for rigid molecules is in fact exact when the interaction 
sites are distributed on the surface of a sphere. Other decoupling schemes 
which have recently been put forward for polymeric systems 
\cite{pago:01,krak:01} are also exact in that limit.

We have considered mixtures of colloids of radius $R_c=150$ nm, 
and of oppositely charged semiflexible polyelectrolyte chains of 
contour length $L=600$ nm and persistence length $l_p=25$ and $50$ nm
in water at room temperature ($e^2/(\varepsilon k_BT)=0.714$ nm).

\newpage
\vspace*{-2.0cm}  
\begin{figure}[h]  
\epsfysize=12cm  
\epsffile{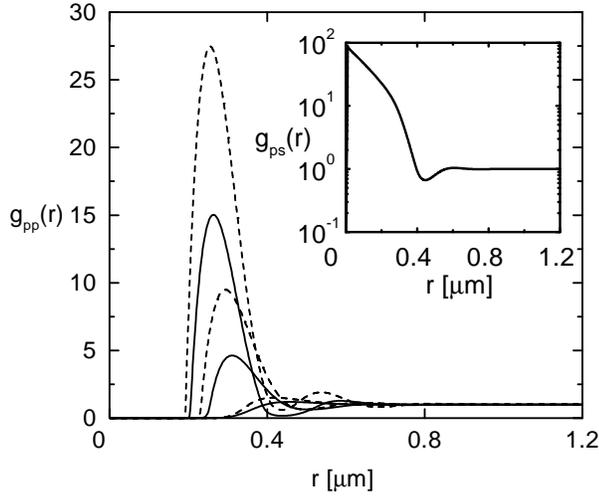}
\vspace*{-3.8cm}  
\caption{Polyelectrolyte pair correlation function $g_{pp}(r)$ 
for a mixture of semiflexible polyelectrolytes (contour length 
$L=600$ nm) and oppositely charged spherical colloids 
(radius $R_c=150$ nm) for various valences $Z_p=-Z_c=200, 400, 600$
(from bottom to top) at \mbox{$\rho_c=0.5$ $\mu$m$^{-3}$}. 
The solid and dashed lines are calculated using the persistence length 
$l_p=100$ nm and $l_p=50$ nm, respectively. The inset displays 
the polyelectrolyte-colloid surface site correlation function 
$g_{ps}(r)$ for $Z_c=-Z_p=400$ and $l_p=100$ nm.}  
\label{fig1}   
\end{figure}

\vspace*{-0.75cm}  
\begin{figure}[h] 
\epsfysize=12cm  
\epsffile{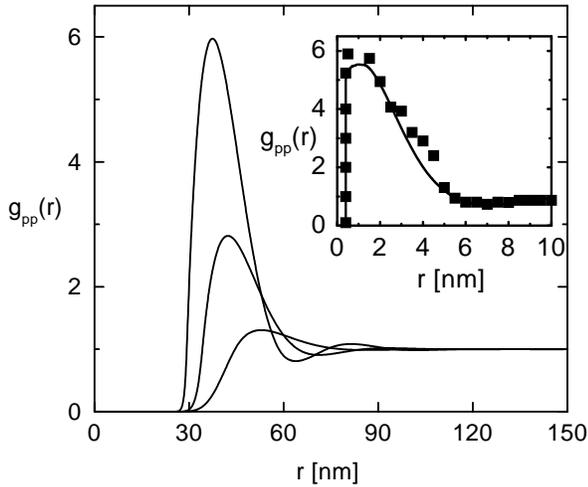}
\vspace*{-3.8cm}
\caption{Polyelectrolyte pair correlation function $g_{pp}(r)$ 
for a mixture of semiflexible polyelectrolytes (contour length 
$L=60$ nm, persistence length $l_p=10$ nm) and oppositely charged 
spherical colloids (radius $R_c=20$ nm) for various valences 
$Z_p=-Z_c=50, 75, 100$ (from bottom to top) at
\mbox{$\rho_c=5\times 10^{-7}$ nm$^{-3}$}. 
The inset displays $g_{pp}(r)$ for a mixture of charged dimers 
($L=1$ nm) and oppositely charged small spherical particles 
($R_c=2$ nm) for $Z_c=-60$ together with computer simulation
data (squares) \protect\cite{resc:00}.} 
\label{fig2}   
\end{figure}

The calculated pair distribution functions 
$g_{pp}(r)$ are shown in Fig.~\ref{fig1} for three values of $Z_c$. 
The interchain correlations are 
seen to increase dramatically with $Z_c$, and to be much more 
pronounced for the more flexible chains.
Qualitatively similar results hold for a mixture involving 
smaller colloidal particles and shorter chains, as shown in  
Fig.~\ref{fig2}. 
The reliability of PRISM is illustrated in the 
inset of that Figure, where the PRISM predictions for $g_{pp}(r)$
is compared to available Monte Carlo data \cite{resc:00} for a 
mixture of charged spheres and oppositely charged dimers.
The unusually large amplitude of the first peak in $g_{pp}(r)$
is the signature of an enhanced polyelectrolyte density near the 
oppositely charged spherical colloids, reminiscent of the well-known 
"counterion condensation" of small ions \cite{mann:69,levi:98}.
This condensation may drive phase separation \cite{resc:00} 
which can be diagnosed by large fluctuations in the colloid density.

\begin{figure}[h] 
\epsfysize=12.0cm  
\epsffile{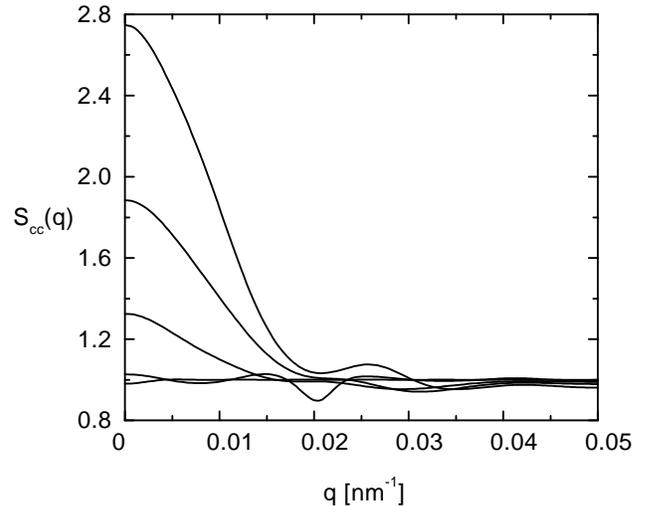}
\vspace*{-3.8cm}
\caption{Colloid centre of mass structure factor $S_{cc}(q)$  
for a mixture of semiflexible polyelectrolytes 
($L=600$ nm, $l_p=50$ nm) and oppositely charged spherical 
colloids ($R_c=150$ nm, $Z_c=1000$, $Z_p=-800$) for various densities 
\mbox{$\rho_c=0.004,\,0.04,\,0.1,\,0.2,\,0.4$ $\mu$m$^{-3}$} 
(from bottom to top).} 
\label{fig3}   
\end{figure}

\begin{figure}[h] 
\epsfysize=12.0cm  
\epsffile{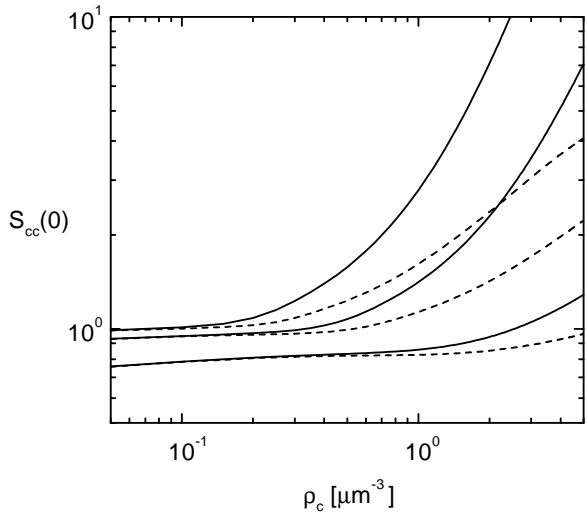}
\vspace*{-3.5cm}
\caption{Colloid centre of mass structure factor at zero 
scattering vector $S_{cc}(0)$  as a function
of colloid density, 
for a mixture of semiflexible polyelectrolytes 
($L=600$ nm) and oppositely charged spherical 
colloids ($R_c=150$ nm) for various valences 
$Z_c=-Z_p=100,\,200,\,300$ (from bottom to top). The solid and dashed 
lines are calculated using the persistence length 
$l_p=25$ nm and \mbox{$l_p=50$ nm}, respectively.} 
\label{fig4}   
\end{figure}

Examples of the colloid centre of mass structure factor 
$S_{cc}(q)$ are shown in Fig.~\ref{fig3}. 
A central peak in the $q \to 0$ limit is 
seen to build up as the colloid density $\rho_c$ increases. The 
enhancement of the density fluctuations, as characterized 
by $S_{cc}(q=0)$, is illustrated in Fig.~\ref{fig4}, as a function 
of $\rho_c$ for several values of $Z_c$.
Although no divergence occurs, the tendency towards segregation is 
clear. In a naive random phase approximation (RPA) picture, values 
of  $S_{cc}(0)>1$ point to an effective long range attraction 
between the colloidal particles. The existence of an attraction 
between the colloidal particles, induced by the oppositely 
charged polyelectrolyte, can be more firmly established by a direct 
inversion of the pair distribution function $g_{ss}(r)$ of surface 
sites on the colloids.

\vspace*{-1.1cm}  

In the limit $\rho_c \to 0$, this effective potential between 
surface sites would coincide with the potential of mean 
force, $v_{ss}(r)=-k_BT \log g_{ss}(r)$, but at finite concentration 
of colloids, an inversion based on an accurate closure relation 
must be used. In other words the inverse problem is solved, 
whereby the pair distribution function $g_{ss}(r)$, as calculated 
from the solution of the multicomponent PRISM equations, is known, 
and the effective potential $v_{ss}(r)$ is extracted from the 
one-component version of the theory, using the corresponding 
one-component OZ and closure relations. Note that the inversion 
does not require an iterative procedure. Similar inversions have 
recently been successfully used to determine effective interactions 
between the centres of mass of interacting polymer coils 
\cite{bolh:01}, and between the interaction sites of rodlike polyelectrolytes 
\cite{hofm:01}.

\begin{figure}[h]  
\epsfysize=12cm  
\epsffile{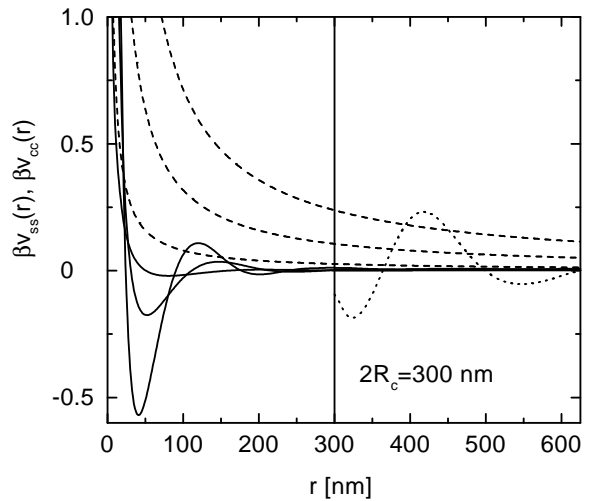}
\vspace*{-3.5cm}  
\caption{Effective potential between surface sites on neighbouring 
colloids (solid line), and the bare Coulomb site-site interaction 
(dashed line) for a mixture of charged spherical colloids (radius $R_c=150$ nm) 
in the presence of  oppositely charged polyelectrolytes 
(contour length $L=600$ nm, persistence length 
$l_p=50$ nm) at \mbox{$\rho_c=5 \mu$m$^{-3}$}.
From top to bottom the solid lines correspond to increasing values 
of valences ($Z_c=-Z_p=200,\,400,\,600$), while the dashed lines correspond 
to decreasing values of $Z_c$. The dotted line represents the effective 
potential $v_{cc}(r)$ between the centres of the colloidal particles as 
derived from an HNC inversion of $g_{cc}(r)$ for $Z_c=-Z_p=200$.}  
\label{fig5}   
\end{figure}

Examples of the effective potentials are shown 
Fig.~\ref{fig5}. As the charge $Z_c e$
increases, an attractive well develops and deepens. Instability towards 
colloid clustering and genuine phase separation is expected to 
occur when the well-depth becomes of the order $k_BT$, so that yet 
stronger Coulomb interaction, i.e., more highly charged polyelectrolyte, 
would be needed to drive the phase separation. This is presently under 
investigation. A comparison with the bare Coulomb interaction (\ref{eq1}) 
clearly demonstrates the enhanced screening effect of the polyelectrolyte 
as the charge increases. The same figure also shows the effective 
potential $v_{cc}(r)$ between the centres of the colloidal particles, 
obtained from an HNC inversion of the centre to centre pair distribution 
\mbox{function $g_{cc}(r)$ for one value of $Z_c$.}

\newpage
\vspace*{-1.1cm}  

\section{Bidisperse clay suspensions}
We now turn to binary mixtures of oppositely charged disc-shaped 
platelets of different radii, $R_1$ and $R_2$ suspended in water, 
in the presence of microscopic co- and counterions. $N_1$ and $N_2$ 
interaction sites are assumed to be uniformly distributed over 
the surfaces of the discs of species $1$ and $2$, and interact 
via the screened Coulomb potential (\ref{eq1}); the hard core 
diameters $d_1$ and $d_2$ determine the effective thickness 
of the platelets. In the limit of a continuous distribution of 
sites, the form factors $\omega_{i}(q)\equiv \omega_{ii}(q)$ 
($1 \le i \le 2$) are easily calculated to be \cite{kutt:00}:
\begin{equation}  \label{eq17}
\omega_i(q)=\frac{2N_i}{(qR_i)^2}\left(1-\frac{J_1(2qR_i)}{qR_i}\right)\,,
\end{equation} 
where $J_1$ denotes the cylindrical Bessel function of first order
and all interaction sites are assumed to be equivalent, i.e., edge 
effects are neglected. This may be a more severe approximation for 
platelets than for spheres (where the equivalence is exact), or 
rigid or semiflexible polyelectrolyte chains, but our previous 
results for the one-component case show that it leads to good 
agreement with available simulation data \cite{harn:01}.
The correlation function matrices ${\bf H}(q)$ and ${\bf C}(q)$
are now $2$ x $2$ matrices, and {\bf W}(q) is diagonal.

\begin{figure}[h]
\vspace*{-1.1cm} 
\begin{center} 
\epsfysize=12cm 
\epsffile{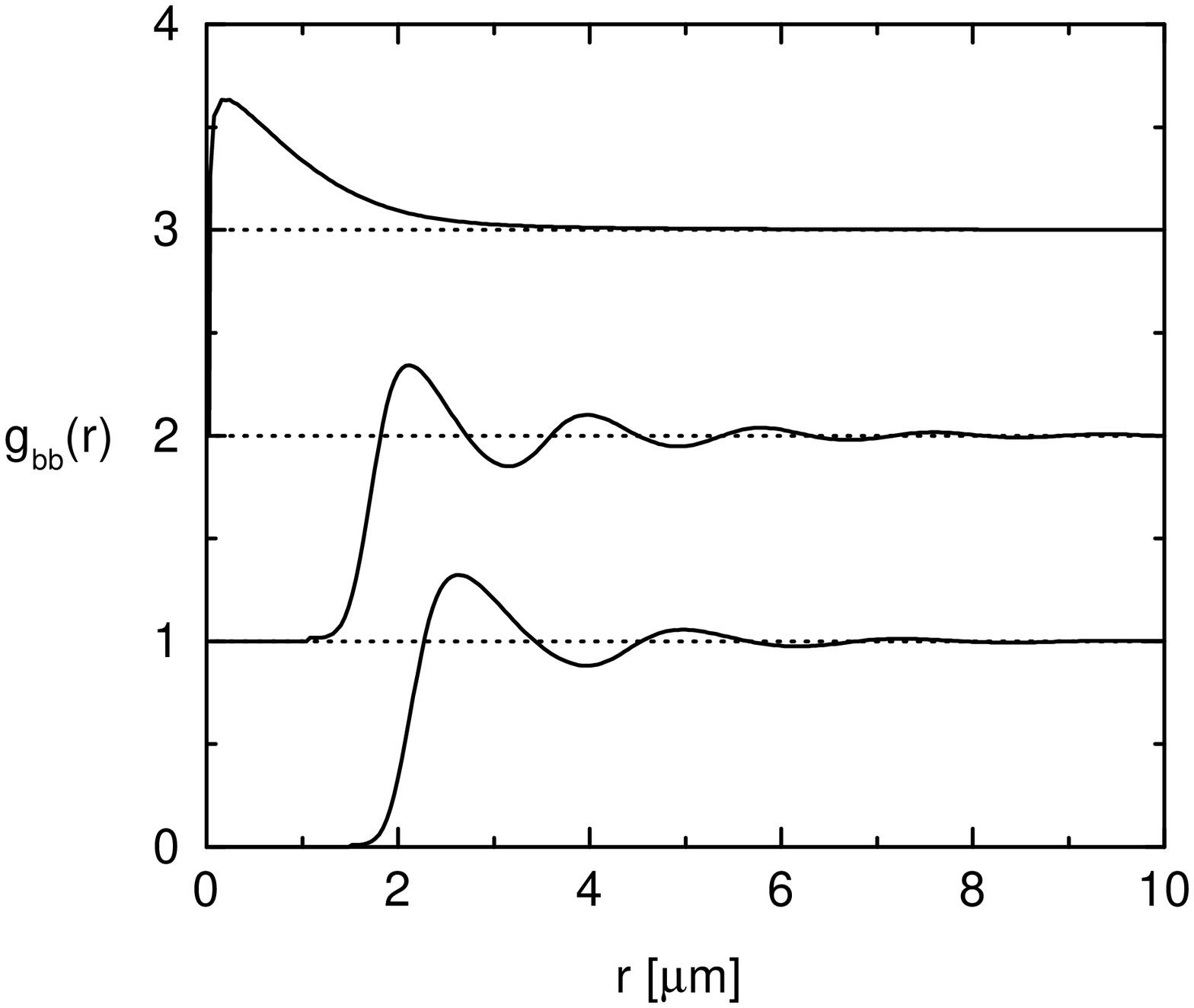} 
\vspace*{-3.5cm} 
\caption{Pair correlation function $g_{bb}(r)$ for a mixture 
of large (subscript $b$) and small, oppositely charged 
(subscript $m$) platelets for various densities:  
$\rho_b=0.183$ $\mu$m$^{-3}$, $\rho_{m}=40$ $\mu$m$^{-3}$  
(top curve); 
$\rho_b=0.074$ $\mu$m$^{-3}$, $\rho_{m}=16$ $\mu$m$^{-3}$  
(middle curve); 
$\rho_b=0.037$ $\mu$m$^{-3}$, $\rho_{m}=8$ $\mu$m$^{-3}$ 
(bottom curve). The upper two curves are shifted.} 
\label{fig7} 
\end{center} 
\end{figure}

\begin{figure}[h]
\begin{center} 
\epsfysize=12cm 
\epsffile{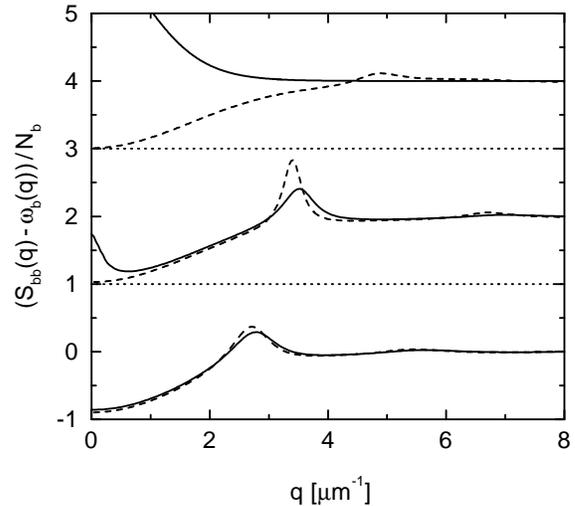} 
\vspace*{-3.5cm} 
\caption{Platelet structure factor for a mixture of large  
(subscript $b$) and small, oppositely charged (subscript $m$)  
platelets for various densities:  
$\rho_b=0.183$ $\mu$m$^{-3}$, $\rho_{m1}=4$ $\mu$m$^{-3}$,  
$\rho_{m2}=40$ $\mu$m$^{-3}$ (top curves); 
$\rho_b=0.074$ $\mu$m$^{-3}$, $\rho_{m1}=1.6$ $\mu$m$^{-3}$,  
$\rho_{m2}=16$ $\mu$m$^{-3}$  (middle curves); 
$\rho_b=0.037$ $\mu$m$^{-3}$, $\rho_{m1}=0.8$ $\mu$m$^{-3}$,  
$\rho_{m2}=8$ $\mu$m$^{-3}$ (bottom curves). The dashed and solid  
lines represent the calculations for $\rho_{m1}$ and $\rho_{m2}$, respectively. 
The upper four  curves are shifted. } 
\label{fig8} 
\end{center} 
\end{figure}

We have solved the PRISM equations for platelet parameters and 
physical conditions appropriate for a water base drilling mud, 
made up of large, negatively charged bentonite platelets, (index $b$), 
and much smaller, positively charged mixed metal hydroxide 
platelets (index $m$). The radii of the two species were taken 
to be $R_1=R_b=10^3$ nm and $R_2=R_m=25$ nm; the effective 
thicknesses were chosen to be $d_b=d_m=10$ nm, mimicking conditions 
where the platelets are not fully dispersed and form stacks. The 
strong screening in water base drilling muds is mainly due to the 
addition of NaOH (pH=12), and this is modelled by adopting an 
inverse screening length, $\kappa_D=1$ nm$^{-1}$. A homogeneous 
interaction site density on both types of platelets is achieved by 
imposing $N_b/N_m=R_b^2/R_m^2=1600$; in practice $N_b=30400$ and 
$N_m=19$, while surface charge densities were fixed at 
\mbox{$\sigma_b=0.09$ e$\,$nm$^{-2}$} and \mbox{$\sigma_m=0.9$ e$\,$nm$^{-2}$}. 
Examples of bentonite pair distribution functions $g_{bb}(r)$ are 
shown in Fig.~\ref{fig7}. At the two lower clay concentrations, 
the position of the first peak scales roughly like $\rho_b^{-1/3}$, 
but at the highest density the peak shifts dramatically towards 
the origin, a behaviour indicative of a collapse of nearly 
parallel pairs of bentonite 
platelets, under the influence of a strong effective attraction, 
induced by the smaller platelets that appear to ''condense'' on the 
surfaces of the bentonite discs. Note that all secondary minima and 
maxima have disappeared from $g_{pp}(r)$ at the highest density. 
The corresponding structure factors $S_{bb}(q)$ (minus the fixed 
intramolecular contribution) are pictured in Fig.~\ref{fig8}, 
together with the results obtained from PRISM for the same 
bentonite densities, but ten times lower densities of 
the smaller platelets. 
The structure factors $S_{bb}(q)$
at the two lower densities show typical 
liquid-like structure, with a main peak at $q=2\pi/D_b$, 
where $D_b\approx \rho_b^{-1/3}$. 
Increasing $\rho_b$ or adding mixed 
metal hydroxide shifts the peak toward higher $q$-values, 
corresponding to a shorter distance  between bentonite platelets.
Increasing $\rho_m$ also leads to an enhancement of the small 
angle ($q\to 0$) value of $S_{bb}(q)$; a clear central peak 
appears at intermediate densities, while at the highest densities 
the amplitude of the central peak appears to diverge, signalling
strong concentration fluctuations indicative of platelet aggregation, 
or spinodal instability reminiscent of the behaviour observed 
in recent computer simulations of mixtures of oppositely charged 
spherical particles \cite{lins:99}. A similar strong small $q$ upturn 
has been observed experimentally for monodisperse clay suspensions in 
the gel phase \cite{morv:94,pign:97,mour:98,kroo:98,saun:99,bonn:99}, 
polyelectrolyte gels \cite{scho:94} and mixtures \cite{norw:96}, 
Poly(styrenesulfonate) ion exchange resins \cite{maar:96}, and  low ionic 
strength polyelectrolyte solutions \cite{foer:90,boue:94,ermi:98}, 
but the exact physical origin of these large-scale fluctuations for 
these systems is still under debate.

\vspace*{-2.7cm} 

\section{Conclusion}
We have illustrated the mechanism of mesoscopic particle aggregation 
induced by oppositely charged polyions on two examples of complex 
two-component dispersions. The partial ''condensation'' of the small 
particles leads to a considerable reduction of the strong Coulomb 
repulsion between the larger colloidal particles, which may go as 
far as an effective short-range attraction between the latter. The 
attraction, and resulting aggregation give rise to a strong enhancement 
of the long wavelength concentration fluctuations of the large 
particles, which is well characterized by the small $q$ limit of the 
corresponding partial structure factor. 

While effective attractions have been widely studied theoretically and 
by simulations in the case of microscopic counterions \cite{hans:00},
PRISM has allowed us to extend such 
studies to cases where the ''counterions'' are themselves complex 
objects (semiflexible polyelectrolyte chains or clay platelets). PRISM
turns out to be an accurate and flexible tool to tackle highly 
asymmetric complex systems of charged particles of various shapes, 
which are beyond the search of present day simulation methodology. 
The PRISM integral equations provide detailed structural information, 
in the form of partial, site-site pair distribution functions and 
structure factors. Thermodynamic properties can then be determined via 
the energy route, or the compressibility route \cite{hans:86}, but the 
structural information is not sufficient to give direct access to the 
equation of state via the viral route, and to the free energy necessary 
to determine phase diagrams, which requires thermodynamic integration.
Work along these lines is in progress, as well as further calculations 
based on model B for the colloid-polyelectrolyte mixture.

\acknowledgements
The authors are grateful to Edo Boek for helpful discussions.

\end{document}